\documentclass[aps,superscriptaddress,preprintnumbers,showpacs]{revtex4}
\usepackage{amssymb}
\usepackage{graphicx}
\usepackage{dcolumn}
\usepackage{bm}

\begin{document}

\newcommand*{\sjtu}{INPAC, Department of Physics, Shanghai Jiao Tong University, Shanghai, China}\affiliation{\sjtu}
\newcommand*{\NTU}{Department of Physics and Center for Theoretical Sciences, National Taiwan University, Taipei, Taiwan}\affiliation{\NTU}

\title{LHC Evidence Of A 126 GeV Higgs Boson\\ From $H \to \gamma \gamma$ With Three And Four Generations}

\author{Gang Guo}\affiliation{\sjtu}
\author{Bo Ren}\affiliation{\sjtu}
\author{Xiao-Gang He}\email{hexg@phys.ntu.edu.tw}\affiliation{\sjtu}\affiliation{\NTU}

\begin{abstract}
Searches for Higgs boson at the LHC have excluded standard model (SM) Higgs boson mass in the range between 127 GeV
to 600 GeV. With a fourth generation, the excluded range is wider. To close the windows
between 114 GeV to 127 GeV, the mode $H \to \gamma\gamma$ plays an important role.
There are evidences that the Higgs boson mass is about 126 GeV from LHC data.
$H\to \gamma\gamma$ can occur at one loop level in the SM. In the SM with three generations (SM3), the dominant contribution is from W boson with some
cancellation from top quark in the loop.  With SM4, the large mass of the fourth generation quarks and charged lepton
cancel the W boson contribution significantly, the decay width is suppressed by a factorin the range of 0.25 $\sim$ 0.55 for the fourth
generation mass in the range of 500 to 1000 GeV. This reduction factor makes $\sigma(pp\to H X)Br(H\to \gamma \gamma)$ for SM4
comparable to that for SM3 for Higgs boson mass in the window allowed mentioned earlier. Using $H \to \gamma \gamma$ alone,
therefore, it is difficult at present to distinguish whether the Higgs boson is from SM3 or SM4. We also comments on
some other detection channels.
\end{abstract}


\maketitle

The LHC has successfully run at 7 TeV and have collected valuable data about Higgs boson. LHC data, up to August 2011, has excluded a large range for
Higgs boson mass in the standard model (SM) with three generations of quark and leptons (SM3) from 141 GeV to 476 GeV at 95\% c.l.~\cite{lhcsm3}.
For the SM with four generations (SM4), a wider range between 120 GeV to 600 GeV is excluded with the same confidence level~\cite{lhcsm4}. Some implications of these data
have been discussed~\cite{implication}. Electroweak precision data prefer
a Higgs boson at about 120 GeV~\cite{electroweak}. Therefore the range allowed by LEP and LHC data
is between 114 GeV to 141 GeV for SM3, and between 114 GeV to 120 GeV for SM4 at 95\% c.l.. The narrower range for SM4 is mainly due to
the fact that a larger production rate for Higgs boson $H$ due to larger $gg\to H$ coupling with SM4 compared with SM3, since the detection of Higgs
boson is mainly by the decay mode $H \to W^+W^-$ which are the same for both SM3 and SM4. The recent data with more statistics, a wider exclusion range for SM3 of 127 GeV to 600 GeV
has been obtained~\cite{LHC-new}.
One should keep in mind that with higher confidence level, the allowed ranges for Higgs boson mass is wider.
The recent LHC data also show indications that the Higgs boson mass is about 126 GeV from both ATLAS and CMS at a confidence level of 2$\sigma$ to 3$\sigma$. It is too early
to draw definite conclusion about the discovery of the Higgs boson at present.

In both ATLAS and CMS analysis the decay channel $H \to \gamma \gamma$ played an important role.
In the range between 114 GeV to 130 GeV, the SM3 $H \to \gamma \gamma$ has reasonably large
branching ratio which has clear signature for detection. Therefore this process plays a crucial role
in discovering the Higgs boson~\cite{LHC-new}. For SM4, the situation is more complicated because on the one hand, the fourth generation quarks are expected to have large masses
which can enhance the $gg \to H$ by about a factor of 9 which makes the Higgs boson production rate larger. On the other hand, the heavy fermions, quarks and charged lepton,
in the fourth generation tend to cancel $W$ boson contribution to $H \to \gamma \gamma$ resulting in a smaller decay branching ratio. Therefore the LHC indication of a Higgs boson with a mass
of 126 GeV may not be simply identified to be that from SM3 using $H \to \gamma \gamma$ alone with the current statistics. More data and more channels should be analyzed to have more precise information.
In the following we provide some details for $H \to \gamma\gamma$ in the SM3 and SM4, and also comment on $H \to ZZ^* \to 4l$ channel.

The decay of the Higgs boson to photons is mediated by $W$ boson and heavy charged fermion at one loop level in the SM, the partial decay width can be
written as~\cite{h22gamma}
\begin{eqnarray}  \label{DH}
\Gamma(H\rightarrow \gamma\gamma)= \frac{G_F \alpha^2 M^3_H}{128 \sqrt{2} \pi^3} \Biggl|\sum_f N_{c} Q^2_f A^H_{1/2}(\tau_f) + A^H_1(\tau_W)\Biggl|^2\;,
\end{eqnarray}
where
\begin{eqnarray}
&&A^H_{1/2}(\tau)= 2\tau^{-2} [\tau+(\tau - 1)f(\tau)]\;,\nonumber\\
&&A^H_W= -\tau^{-2} [2 \tau^2 +3\tau +3(2\tau-1)f(\tau)]\;,
\end{eqnarray}
are for contributions from spin-1/2 and spin $1$ particles in the loop, respectively.
The function $f(\tau)$ is defined by
\begin{eqnarray}
f(\tau)=\left\{\begin{array}{ll}\arcsin^2\sqrt{\tau} &\hspace{2em} \tau\geq1\\
-\frac{1}{4}\Biggl[\log\frac{1+\sqrt{1-\tau}}{1-\sqrt{1-\tau}}-i\tau\Biggl]^2 &\hspace{2em} \tau<1\end{array}\right.
\end{eqnarray}
The parameter $\tau_i$ is defined by $\tau_i=M^2_H/4M^2_i$ with $i=f, W$ the corresponding heavy fermions and W boson.

The QCD corrections of the quark contribution to the two-photon Higgs decay amplitude can be parameterized as
\begin{eqnarray} \label{qcd}
A^H_{1/2}(\tau_Q) = A^H_{1/2}(\tau_Q)|_{LO}\Biggl[1+\frac{\alpha_s}{\pi}C_H(\tau_Q)\Biggl].
\end{eqnarray}
In principle, the scale in $\alpha_s$ is, typically, taken to be of order $M_H$.
Two-loop corrections in the on-shell renormalization scheme for $C_H(\tau_Q)$ show that this correction is about 5\% \cite{qcd}.

The electroweak corrections of $\mathcal{O}(G_FM^2_t)$ have also been evaluated. This part of the correction arises from all diagrams, which contain a
top quark coupling to a Higgs particle or would-be Goldstone boson. The final expression results in a re-scaling factor to the top quark
loop amplitude, given by \cite{Liao}
\begin{eqnarray} \label{cn}
A^H_t(\tau_t)\rightarrow A^H_t(\tau_t)\times \Biggl[1-\frac{3}{4e^2_t} (4e_te_b +5 -\frac{14}{3}e^2_t)\frac{G_F M^2_t}{8\sqrt{2}\pi^2} \Biggl].
\end{eqnarray}
where $e_{t,b}$ are the electric charges of the top and bottom quarks, The effect is enhancement of the photonic decay width by less than $1\%$,
so that these corrections are negligible at most circumstance. For SM4, one need to add contributions from the fourth generation quarks, $t'$ and $b'$, and
charged lepton $l'$.

The leading contribution to the decay of the Higgs boson into two gluons is generated at the one loop level by heavy quarks,
with the main contribution coming from top quarks and a small contribution from bottom quarks for SM3. At the leading order,
the partial decay width reads \cite{Wilczek}
\begin{eqnarray} \label{gg}
\Gamma(H\rightarrow gg)= \frac{G_F \alpha^2_s M^3_H}{36 \sqrt{2}\pi^3}  \Biggl|\frac{3}{4} \sum_Q A^H_{1/2}(\tau_Q) \Biggl|^2.
\end{eqnarray}
Here $\tau_Q= M^2_H/4m^2_Q$ for a heavy quark $Q$. The form factor $A^H_{1/2}(\tau_Q)$ is similar
to that for heavy fermion contribution to the $H \rightarrow \gamma\gamma$ case.

The calculation of the next leading order (NLO) QCD correction in the full massive case has been performed in Ref. \cite{Spira} where
the rather complicated analytical expressions can be found. The total correction can be cast into the form
\begin{eqnarray} \label{ggg}
\Gamma(H \rightarrow gg (g), g q \bar{q})= \Gamma_{LO} (H \rightarrow gg) \Biggl[1 + E_H(\tau_Q) \frac{\alpha_s}{\pi} \Biggl].
\end{eqnarray}
and one obtains for the correction factor
\begin{eqnarray}
E_H(\tau_Q) = \frac{95}{4} -\frac{7}{6}N_f + \frac{33-2N_f}{6}\log\frac{\mu^2}{M^2_H}+\Delta E_H(\tau_Q).
\end{eqnarray}
where $\mu$ is the renormalization point which is about the Higgs mass and defines the scale of $\alpha_s$. The first three terms survive in the limit of
large loop masses while $\Delta E_H$ vanishes in this limit \cite{Inami}. The QCD radiative corrections turn out to be quite important,
nearly doubling the gluonic partial decay width. In the mass range $M_H\leq 2M_W$, the QCD corrections leading to an increase of the partial
width by $~70\%$.

The electroweak corrections of $\mathcal{O} (G_F M^2_t)$ to the gluonic decay width, which are mediated by virtual top quarks, have also been
calculated. The final result leads to a simple rescaling of top quark amplitude, given by \cite{Djouadi}
\begin{eqnarray}
A^H_t (\tau_t) \rightarrow A^H_t (\tau_t) \times (1+ \frac{G_f \sqrt{2}
} {32 \pi^2} M_t^2 ).
\end{eqnarray}
They enhance the gluonic decay width by about $0.3\%$ which is negligible. In the case of a 4th family of heavy quarks with degenerate masses, each heavy quark amplitude should be re-scaled in the form \cite{Djouadi}
\begin{eqnarray}
A^H_i (\tau_i) \rightarrow A^H_i (\tau_i) \times (1- \frac{G_f \sqrt{2}
} {8 \pi^2} M_i^2 ),      ~~~~~~ \textit{where} ~~~i = t', b'.
\end{eqnarray}

With the above information, we can now compare the SM3 and SM4 predictions~\cite{calculations} for $H \to \gamma\gamma$ and also $\sigma(SM)=\sigma(pp \to H X)_{SM}Br(\gamma \gamma X)_{SM}$.
The cross section $\sigma(SM)$ gauges how many Higgs boson are produced and subsequently decay into $\gamma\gamma$. It is proportional to
$\Gamma(H \to gg)Br(H \to \gamma\gamma)$. For SM4 predictions, one needs to know the fourth generation fermion masses. The current searches at the LHC have set
the fourth generation masses to be larger than 490 GeV~\cite{lhc4mass}. We will study the effects of the fourth generation mass above the current limit to 1 TeV for illustration.
Our results are shown in Figs.\,1 to 3.

In Fig.\,1, we show the ratio $\Gamma(H\to \gamma\gamma)_{SM_4}/\Gamma(H\to \gamma\gamma)_{SM_3}$. The partial width in SM3 for Higgs boson mass in the
range between 114 GeV to 130 GeV is in the range 0.003 GeV to 0.005 GeV. We see that the SM4 is suppressed, but the cancellation is not complete. The suppression factor is above 0.25 $\sim$ 0.55. This suppression does not mean that the SM4 Higgs is more difficult to search at the LHC compared with SM3 using $H \to \gamma\gamma$. One should also consider the productions for $H$ at the LHC.

\begin{figure}[t!]
\includegraphics[width=4 in]{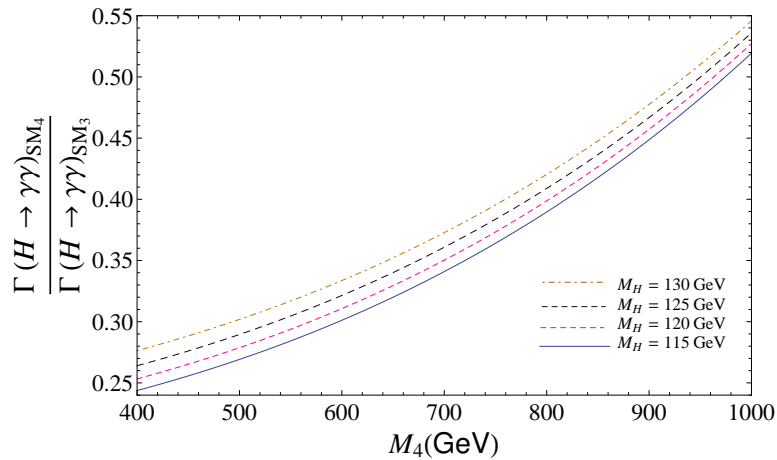}
\caption{\label{feynman1} $\Gamma(H\to \gamma\gamma)_{SM4}/\Gamma(H\to \gamma\gamma)_{SM3}$ vs. fourth generation fermion mass for several different Higgs boson masses.}
\end{figure}

The main mechanism for Higgs productions at the LHC pp collider is $gg \to H$. Compared with SM3, the SM4 is approximately enhance be a factor of 9. The enhanced production
of Higgs boson for SM4 can over take the suppression for $H \to \gamma\gamma$ and end up with comparable $\sigma(SM_3)$ and $\sigma(SM_4)$. To obtain $\sigma(SM) = \sigma(pp \to H X)_{SM}Br(H \to \gamma \gamma)_{SM}$, one needs to obtain the total Higgs boson decay width $\Gamma(\mbox{total})_{SM}$ for SM3 and SM4 to make a detailed comparison. We show in Fig.\,2. the SM4 branching ratios and the ratio of the total branching ratios. In the Higgs boson mass range between 114 GeV to 130 GeV, the one loop contribution for $H \to gg$ is non-negligible for SM3 as can be see from Fig.\,2a. For SM4, $H\to gg$ contributes roughly one third of the total decay width. This leads to $\Gamma(\mbox{total})_{SM_4}/\Gamma(\mbox{total})_{SM_3} \sim 1.5 \sim 1.6$ as can be seen from Fig.\,2b. The enhanced total Higgs boson decay width for SM4 reduces the branching ratio for $H\to \gamma\gamma$.

\begin{figure}[t!]
\includegraphics[width=3.1 in]{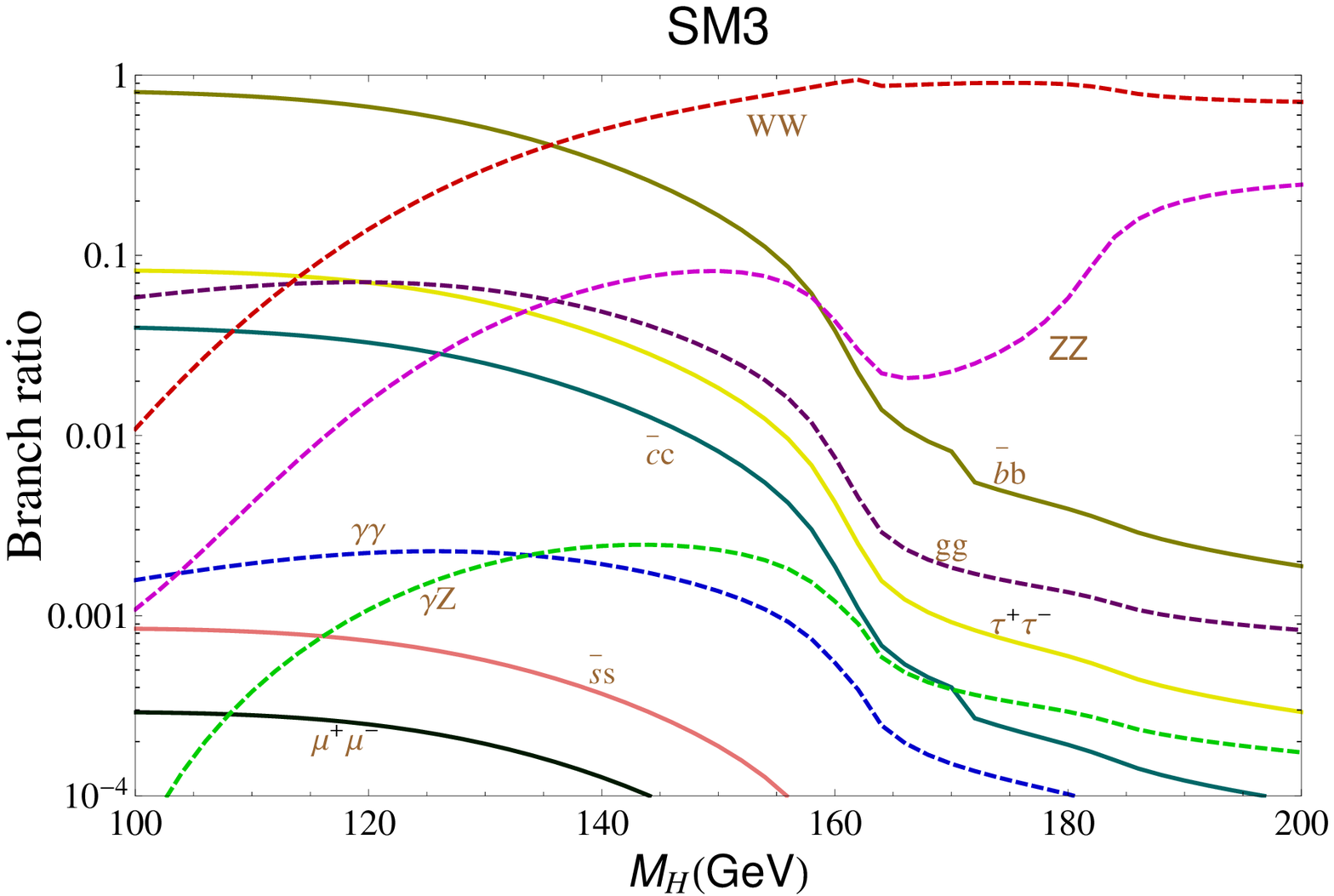}\hspace{0.5 cm} \includegraphics[width=3.3 in]{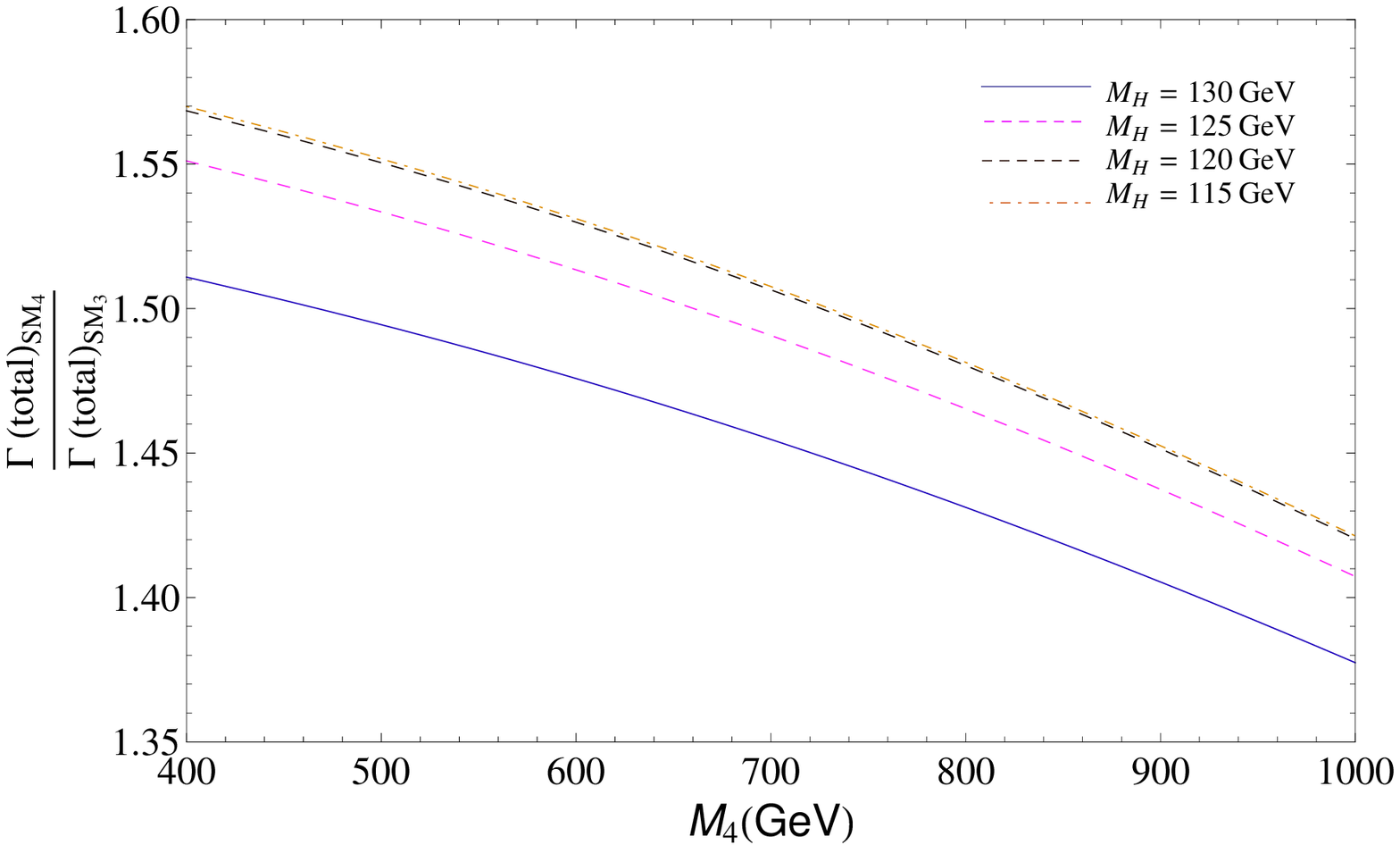}
\caption{\label{feynman2} a). (left) SM3 Higgs boson decay branching ratio for different processes. b). (right) The ratio for  $\Gamma(\mbox{total})_{SM_4}/\Gamma(\mbox{total})_{SM_3}$ vs. fourth generation fermion mass for different Higgs boson masses.}
\end{figure}

Combining these results, we finally obtained the ratio $\sigma(\gamma \gamma, SM_4)/\sigma(\gamma\gamma, SM_3)$ in Fig.3. We see that the ratio of event numbers of $H \to \gamma\gamma$ for SM3 and SM4 is only about 1.5 to 2.5. This in principle can be used to distinguish Higgs boson from SM3 from that from SM4. However, at the present the statistics is not good enough to claim that the indication of Higgs boson of a mass around 126 GeV to be from SM3. More data is needed.

\begin{figure}[t!]
\includegraphics[width=4 in]{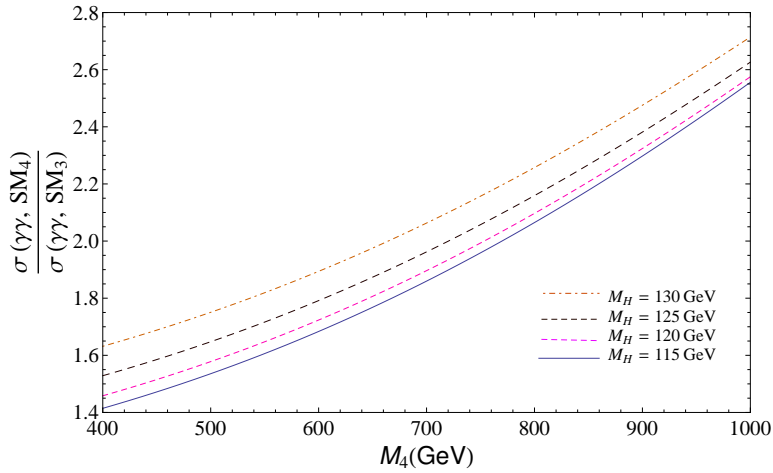}
\caption{\label{feynman3} $\sigma(\gamma\gamma, SM_4)/\sigma(\gamma\gamma, SM_3)$ vs. fourth generation fermions mass for several Higgs boson masses.}
\end{figure}

Besides having more statistics for $H \to \gamma\gamma$, one can also combine other methods to further distinguish whether the Higgs boson is from SM3 or SM4. Compared with SM3, SM4 has larger $H \to gg$ which will enhance the hadronic decay modes for $\Gamma(\mbox{total})_{SM_4}$. Therefore the total Higgs boson decay width can be used to distinguish the models. This may be achieved with more statistics. A more direct one can be from $H \to ZZ^*\to 4l$.
This channel has also been analyzed at ATLAS and CMD~\cite{LHC-new} which support the Higgs boson evidence with the same mass from $H\to \gamma \gamma$. Again the statistics at present still cannot claim the discovery of Higgs boson.
In SM4, the enhancement for $H\to Z Z^* \to 4l$ is larger compared with that for $H\to \gamma\gamma$. This is because that the decay width for $H\to ZZ^*$ is not changed for SM3 and SM4 to the leading order. One would naively think that the event rate in this case should be enhanced by a factor of 9 for SM4, the same enhancement factor for $\Gamma(H\to gg)$. However, one should also consider the enhancement to the total decay width. It is $\sigma(ZZ^*, SM) = \sigma(pp\to H X)Br(H\to ZZ^*)$ which determines the event rate. In Fig.\,4, we show the ratio $\sigma(ZZ^*, SM_4)/\sigma(ZZ^*, SM_3)$. We see that there is a factor of 5 only enhancement for SM4 compared with SM3.
Currently the exclusion range for SM4 Higgs boson is from 120 GeV to 600 GeV at 95\% c.l.. At higher confidence level, a 125 GeV cannot be conclusively ruled out at present. To completely close the window more statistics is needed. The LHC data from next year can provide more definitive conclusions about the existence of Higgs boson, and whether it is from SM3 or SM4 if it is found.
\\

\begin{figure}[t!]
\includegraphics[width=4 in]{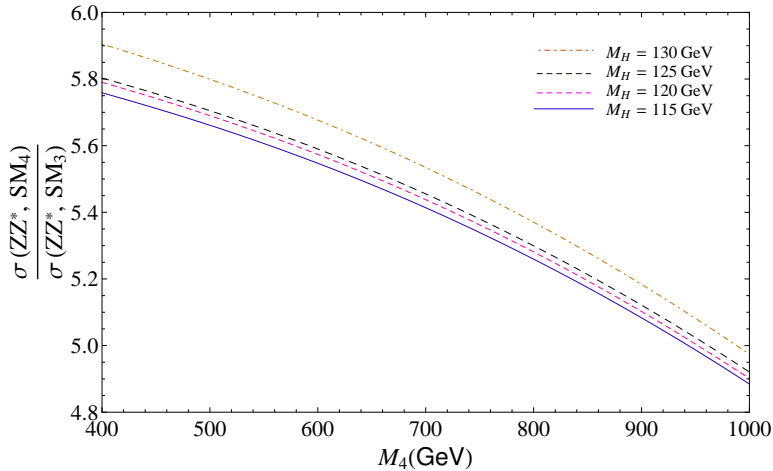}
\caption{\label{feynman4} $\sigma(ZZ^*, SM_4)/\sigma(ZZ^*, SM_3)$ vs. fourth generation fermions mass for several Higgs boson masses.}
\end{figure}

\noindent
{\bf Acknowledgement}:
This work was supported in part by NSC, NCTS, and NNSF.


\newpage

\begin{thebibliography}{99}

\bibitem{lhcsm3}ATLAS Collaboration, Report No. ATLAS-CONF-2011-135, 2011,\\
https://twiki.cern.ch/twiki/bin/view/AtlasPublic/AtlasResultsEPS2011;\\
CMS Collaboration, Report No. CMS PAS HIG-11-022, 2011,
http://cms.web.cern.ch/cms/News/2011/LP11.

\bibitem{lhcsm4} 
CDF Collaboration. [arXiv:1107.3875 [hep-ex]];
Andrey Koryton, CMS Collaboration, EPS-HEP 2011, July 21 -27, 2011; Grenoble, RhoneAlpes France.

\bibitem{implication}
W.Y.Keung, P.Schwaller,
JHEP {\bf 1106} 054 (2011). [arXiv:1103.3765 [hep-ph]];
C.~Englert, T.~Plehn, D.~Zerwas and P.~M.~Zerwas,
  Phys.\ Lett.\  B {\bf 703}, 298 (2011)
  [arXiv:1106.3097 [hep-ph]];
X.-G.He and G. Valencia
[arXiv:1108.0222 [hep-ph]];
M.~Raidal and A.~Strumia,
  Phys.\ Rev.\  D {\bf 84}, 077701 (2011)
  [arXiv:1108.4903 [hep-ph]];
Y.~Mambrini,
  arXiv:1108.0671 [hep-ph];
  R.~Foot, A.~Kobakhidze and R.~R.~Volkas,
  Phys.\ Rev.\  D {\bf 84}, 095032 (2011)
  [arXiv:1109.0919 [hep-ph]];
X.~G.~He and J.~Tandean,
  Phys.\ Rev.\  D {\bf 84}, 075018 (2011)
  [arXiv:1109.1277 [hep-ph]];
D.~Borah,
  arXiv:1109.3363 [hep-ph];
  E.~Ma,
  arXiv:1109.4177 [hep-ph];
  Y.~Kajiyama, H.~Okada and T.~Toma,
  arXiv:1109.2722 [hep-ph];
  T.~Cohen, J.~Kearney, A.~Pierce and D.~Tucker-Smith,
  arXiv:1109.2604 [hep-ph];
  I.~Low, P.~Schwaller, G.~Shaughnessy and C.~E.~M.~Wagner,
  arXiv:1110.4405 [hep-ph];
  M.~Pospelov and A.~Ritz,
  Phys.\ Rev.\  D {\bf 84}, 113001 (2011)
  [arXiv:1109.4872 [hep-ph]];
  C.~Englert, J.~Jaeckel, E.~Re and M.~Spannowsky,
  arXiv:1111.1719 [hep-ph];
  Y.~Mambrini,
  arXiv:1112.0011 [hep-ph];
  X.~Chu, T.~Hambye and M.~H.~G.~Tytgat,
  arXiv:1112.0493 [hep-ph];
  Bogdan A. Dobrescu, Graham D. Kribs, Adam Martin, arXiv:1112.2208 [hep-ph];
  C.~Englert, T.~Plehn, M.~Rauch, D.~Zerwas and P.~M.~Zerwas,
  arXiv:1112.3007;
 O.~Lebedev, H.~M.~Lee and Y.~Mambrini,
  arXiv:1111.4482 [hep-ph].


\bibitem{electroweak}
M. Baak, M. Goebel, J. Haller, A. Hoecker, D. Ludwig, K. Moenig, M. Schott, J. Stelzer.
arXiv:1107.0975 [hep-ph];
Martin Goebel, for the Gfitter Group. [arXiv:1012.1331 [hep-ph]].

\bibitem{LHC-new} F. Gianotti, CERN public Seminar `` Update on the standard model HIggs searches in ATLAS'', December, 2011;
 G. Tonelli, CERN public Seminar `` Update on the standard model HIggs searches in ATLAS'', December, 2011.

\bibitem{h22gamma}A.I. Vainshtein, M.B. Voloshin, V.I. Zakharov and M.A. Shifman, Sov. J. Nucl. Phys.
30, 711 (1979).


\bibitem{qcd}
S.Dawson and R.P.Kauffman,Phys. Rev. D {\bf 47}, 1264 (1993);
M. Steinhauser,Preprint MPI/phT/96-130. [arXiv:9612395 [hep-ph]];
J.Fleischer,O.V.Tarasov and V.O.Tarasov,Phys. Lett. B {\bf 584}, 294 (2004). [arXiv:0401090v2 [hep-ph]].

\bibitem{Liao}
Y. Liao, X. Li,
Phys. Lett. B {\bf 396}, 225 (1997). [arXiv:9605310 [hep-ph]].

\bibitem{Wilczek}
F. Wilczek, Phys. Rev. Lett. {\bf 39}, 1304 (1977); J. Ellis, M. Gaillard, D. Nanopoulos and
C. Sachrajda, Phys. Lett. B {\bf 83}, 339 (1979); T. Rizzo, Phys. Rev. D {\bf 22}, 178(1980);
H.Georgi, S.Glashow, M.Machacek and D.Nanopoulos, Phys. Rev. Lett. {\bf 40}, 692 (1978).

\bibitem{Spira}
M. Spira, A. Djouadi, D. Graudenz and P.M. Zerwas, Nucl. Phys. B {\bf 453}, 17 (1995).

\bibitem{Inami}
T. Inami, T. Kubota and Y. Okada, Z. Phys. C {\bf 18}, 69 (1983) ;
A. Djouadi, M. Spira and P.M. Zerwas, Phys. Lett. B {\bf 264}, 440 (1991);
M. Spira, A. Djouadi, D. Graudenz and P.M. Zerwas, Phys. Lett. B {\bf 318}, 347 (1993);
S. Dawson and R.P. Kauffman, Phys. Rev. D {\bf 49}, 2298 (1994).

\bibitem{Djouadi}
A. Djouadi and P. Gambino, Phys. Rev. Lett. {\bf 73}, 2528 (1994);
K.G. Chetyrkin, B.A. Kniehl and M. Steinhauser, Phys. Rev. Lett. {\bf 78}, 594 (1997)
and Nucl. Phys. B {\bf 490}, 19 (1997).

\bibitem{lhc4mass} The CMS Collaboration, CMS PAS EXO-11-054.

\bibitem{calculations}
Q.Li, M.Spira, J.Gao and C.S.Li, Phys. Rev. D {\bf 83}, 094018 (2011);
X. Ruan and Z. Zhang, arXiv:1105.1634 [hep-ph]; G.~D.~Kribs, T.~Plehn, M.~Spannowsky and T.~M.~P.~Tait,
Phys.\ Rev.\ D {\bf 76}, 075016 (2007)  [arXiv:0706.3718 [hep-ph]].



\end{thebibliography}
\end{document}